\documentclass[prl,twocolumn,showpacs]{revtex4}
\usepackage{graphicx,amsmath,amssymb}

\newcommand{\sqrtbj}{v_{\scriptscriptstyle\!B}^{\scriptscriptstyle 1/2}}

\begin{document}

\title{Ionic phase transitions in non-ideal systems}

\author{Kyle J.~Welch}

\author{Fred Gittes}

\affiliation{Department of Physics \& Astronomy, Washington State University,
Pullman, WA 99164-2814 }

\date{\today}

\begin{abstract}
We construct an explicitly solvable Landau mean-field theory for
volume phase transitions of confined or fixed ions driven by
relative concentrations of divalent and monovalent counterions.
Such phase transitions have been widely studied in ionic gels,
where the mechanism relies on self-attraction or elasticity of a
network.  We find here that non-ideal behavior of ions in aqueous
solution can in theory drive phase transitions without a
self-attracting or elastic network.  We represent non-ideality by a
Debye-H\"uckel-like power-law activity, or correlation free energy,
and retain a
mechanical self-repulsion to avoid runaway collapse due to the
non-ideal term.  Within this model we find a continuous line of
gas-liquid-type critical points, connecting a purely monovalent,
divalent-sensitive critical point at one extreme with a divalent,
monovalent-sensitive critical point at the other.  An alternative
representation of the Landau functional handles the second case.
We include a formula for electrical potential, which may be a
convenient proxy for critically varying volume.  Our relatively
simple mean-field formulation may facilitate explorations of
tunable critical sensitivity in areas such as ion detection
technology and biological osmotic control.

\end{abstract}

\pacs{
82.60.-s,
05.70.Jk,
87.15.Zg,
64.70.Nd,
82.60.Fa,
64.70.fh
}

\maketitle

In charged polymer networks, phase transitions were observed and
explained theoretically some time ago by Tanaka and co-workers
\cite{Tanaka-1977,Tanaka-1978,Tanaka-1980,Tanaka-1985}.
In experiments, gels can exhibit ionically-driven collapse or expansion
as the osmotic pressure responds to slight changes in external ion or
composition.  
Volume phase transitions
have long been discussed with an eye to applications 
\cite{Tanaka-1982,Jackson-1997} and as candidate mechanisms
for essential biophysical processes \cite{Tasaki-Verdugo}.
In previous models network self-attraction and restoring elasticity 
has played a crucial role, invoking 
specific elastic properties of polymer
networks \cite{Flory-1953}.

Here we seek to recast the theory of ionic phase transitions into a simple
and mathematically solvable mean-field formulation, including (in simplified
form) the non-ideality that is present in all ionic solutions.  We find that
in principle a self-attracting network is not necessary for a discontinuous
phase transition in the presence of a simplified Debye-H\"uckel-type non-ideality,
because non-ideality itself acts as an effective self-attraction that mediates
an ionically driven phase transition of the gas-liquid type.  Our model
requires us, however, to attribute a self-repulsion to our fixed-ion system
to avoid runaway collapse due to the non-ideal term.

Our simplified formulation of ionic transitions may, in practical
terms, provide a tunable critical sensitivity to ion valence or
concentration.  This suggests, for example, a mechanism for
biophysical cellular functions such as homeostasis.  The theory
may also lend itself to engineering applications involving ion
detection.  For these purposes electric potential $\Phi$,
whose behavior is predicted, might often be a more convenient
dependent parameter than volume in applications to ion detection
or in biophysical roles for critical ionic sensitivity.

Consider a population of $N_0$ ions each of charge $q_0$ that may be
bound to a mechanical structure, such as a
polymer network, or otherwise confined within a volume 
permeable to counterions.  These are the conditions for Donnan
equilibrium \cite{Friedman-Starzak} in which
osmotic pressure of excess counterions goes
hand-in-hand with an internal voltage $\Phi$, relative to
outside, with the same sign as $q_0$.
Even for $\Phi\ne 0$,
neutrality holds to a good approximation
whenever the voltage drop $\Phi$ occurs only at the boundary,
or more generally if $N_0 \gg
C\Phi/q_0$, where $C$ is the system capacitance.

The immobile ions, confined within a variable volume $V$,
have concentration ${c} = N_0/V = 1/v$. 
Monovalent and divalent counterions with charges
$q_a=-q_0$ and $q_b=-2q_0$ are introduced at external concentrations $a$ and $b$,
and have within $V$ the concentrations ${a'}=N_a/V$
and ${b'}=N_b/V$. 
Their free diffusion in and out
is controlled by the the electrical potential $\Phi$.
We introduce nonideality of mobile ions by way of 
a Debye-H\"uckel interaction or correlation free energy
\cite{Landau}
\begin{align}
    {F}_{\scriptscriptstyle{\text{DH}}}
    &\;=\;
    -
    \sqrtbj\,{k_{\text{\tiny B}}T}\, 
    V \big( \underset{\alpha}{\textstyle\sum} z^2_{\alpha} {c}_\alpha \big)^{3/2}
\end{align}
where $q_\alpha = z_\alpha e$.
The essential features of 
${F}_{\scriptscriptstyle{\text{DH}}}$ are that it is negative and contains mobile-ion
concentrations raised to the $3/2$ power.  
The constant
$v_{\scriptscriptstyle B}$ is proportional to 
the cube of Bjerrum length.
In standard aqueous conditions,
\begin{gather}
    \frac{1}{v_{\scriptscriptstyle B}}
    \;=\;
    (12\pi)^2
    \Big(
    \frac{\epsilon\,{k_{\text{\tiny B}}T}}{e^2}
    \Big)^{3}
    \;\approx\;
    3.4 {\,\mathrm{molar}}
    .
\end{gather}
To obtain closed-form solutions,
we omit confined ions from the
sum, and replace the factor $( {a'} + 4{b'}
)^{3/2}$ by $( {a'} + 2{b'} )^{3/2}$,
which will equal ${c}^{3/2}$ once neutrality is imposed.  Thus
we employ the Debye-H\"uckel-like interaction term
\begin{align}
    {F}_\text{int}
    &\;=\;
    -
    \sqrtbj\,{k_{\text{\tiny B}}T}\, 
    V 
    ( {a'} + 2{b'} )^{3/2}
    .
\end{align}
With ${a'}=N_a/V$ and ${b'}=N_b/V$ and
taking derivatives,
\begin{align}
    -{ \partial{F}_\text{int} /\partial V}
    &=
    P_\text{int}
    =
    -\tfrac{1}{2}
    \sqrtbj\,{k_{\text{\tiny B}}T}\, 
    ( {a'} + 2{b'} )^{3/2}
    \\
    { \partial{F}_\text{int} /\partial N_a}
    &=
    \,{k_{\text{\tiny B}}T} 
    \ln\gamma_a
    =
    -\tfrac{3}{2}
    \,{k_{\text{\tiny B}}T} 
    v_{\scriptscriptstyle B}^{\scriptscriptstyle 1/2}( {a'} + 2{b'} )^{1/2}
    \\
    { \partial{F}_\text{int} /\partial N_b}
    &=
    \,{k_{\text{\tiny B}}T} 
    \ln\gamma_b
    =
    -3
    \sqrtbj{k_{\text{\tiny B}}T}\, 
    ( {a'} + 2{b'} )^{1/2}
\end{align}
Here
$\gamma_a$
and $\gamma_b$
are
activity
coefficients \cite{Moore-Dickerson} for the monovalent and divalent ions:
$\mu_a = \mu_a^0 + {k_{\text{\tiny B}}T}\ln\gamma_a a$.
The proportionality of $\ln\gamma$ to the square root of ionic concentration 
represents Debye-H\"uckel behavior, a theory appropriate to
low ionic strength \cite{Resibois-Horvath}.
The free energy, with $\Phi$ externally controlled, is
\begin{multline}
    {F}(N_a,N_b,V,\Phi)
    =
    {F}_\text{int}(N_a,N_b,V) + (N_0\!-\! N_a\! -\! 2N_b )q_0\Phi
    \\
    +{F}_0(V)
    \, + \, 
    {k_{\text{\tiny B}}T} 
    \Big[
    N_a \ln\frac{N_a}{c_0 eV}
    +N_b \ln\frac{N_b}{c_0 eV}
    \Big]
\end{multline}
(with $e$ the natural log base).
Here $F_0(V)$ describes mechanical constraints on the immobile ions,
such as a polymer network carrying the fixed ions or a membrane
containing them.  The logarithmic terms are the ideal free energy of
the mobile ions, and the concentration scale $c_0$ will cancel out.
The mechanical contribution to pressure is
$-\partial F_0 /\partial V = P_0$ (equivalently a function of $v$ or $c$) and
the overall pressure and the
chemical potentials 
$\mu_a=\partial F/\partial N_a$
and 
$\mu_b=\partial F/\partial N_b$
are
\begin{align}
    P
    &\;=\;
    P_0(v)
    +({a'}+{b'}){k_{\text{\tiny B}}T}
    + P_\text{int}
    \\
    \mu_a
    &\;=\;
    -q_0\Phi + {k_{\text{\tiny B}}T}\ln(\gamma_a{a'}/c_0)
    \\
    \mu_b
    &\;=\;
    -2q_0\Phi + {k_{\text{\tiny B}}T}\ln(\gamma_b{b'}/c_0)
\end{align}
If we set 
$\mu_a = {k_{\text{\tiny B}}T}\ln(a/c_0)$
and $\mu_b = {k_{\text{\tiny B}}T}\ln(b/c_0)$, $a$ and $b$
become effective concentrations, referred to an 
ideal external solution.
To fix $\Phi$,
we impose neutrality,
${c}={a'}+2{b'}$, and
introduce a dimensionless potential $\phi= q_0\Phi/{k_{\text{\tiny B}}T}$
which will always be positive
(since $\Phi$ will always be of the same sign as $q_0$).
We have altogether
\begin{gather}
    P
    \;=\;
    P_0(N_0/{c})
    +\tfrac{1}{2}({a'}+{c}){k_{\text{\tiny B}}T}
    -\tfrac{1}{2}
    \sqrtbj\,
    {c}^{3/2}{k_{\text{\tiny B}}T}
    \\
    \phi \;=\;
    \ln(a'/a)
    -\tfrac{3}{2}
    \sqrtbj\, 
    {c}^{1/2}
    \\
    2\phi \;=\;
    \ln(b'/b)
    -3 \sqrtbj\, 
    {c}^{1/2}
\end{gather}
which, together with ${c}={a'}+2{b'} = N_0/v$,
we will solve to find $P(v)$.
Combining the equations involving $\phi$,
putting $2{b'}+{a'}-{c}= 0$ for neutrality, 
and solving yields 
\begin{align}
    {a'}
    &\;=\;
    (a^2/4b)
    \big
    [
    \big(1+ (8b/a^2){c}\big)^{1/2} - 1
    \big]
\end{align}
We define the dimensionless divalent parameter
\begin{align}
    \beta
    &\;=\;
    \frac{8b}{v_{\scriptscriptstyle B}a^2}
    .
\end{align}
To motivate use of $\beta$,
imagine $n$ divalent ions $B$ cooperatively exchanging with
$2n$ monovalent ions $A$, according to
$CA_{2n} + nB \leftrightharpoons CB_{n} + 2nA$.
Equilibrium is then
\begin{align}
    \frac{[CB_{n}]}{[CA_{2n}]}
    &\;=\;
    K\frac{[B]^n}{[A]^{2n}}
    \;\propto\;
    \bigg(
    \frac{b}{a^2}
    \bigg)^n
\end{align}
For large $n$, 
the discontinuous transition 
$CB_{n}\rightarrow CA_{2n}$ occurs at 
a fixed value of $b/a^2$.

\begin{figure}
\includegraphics[width=1.0\columnwidth]
{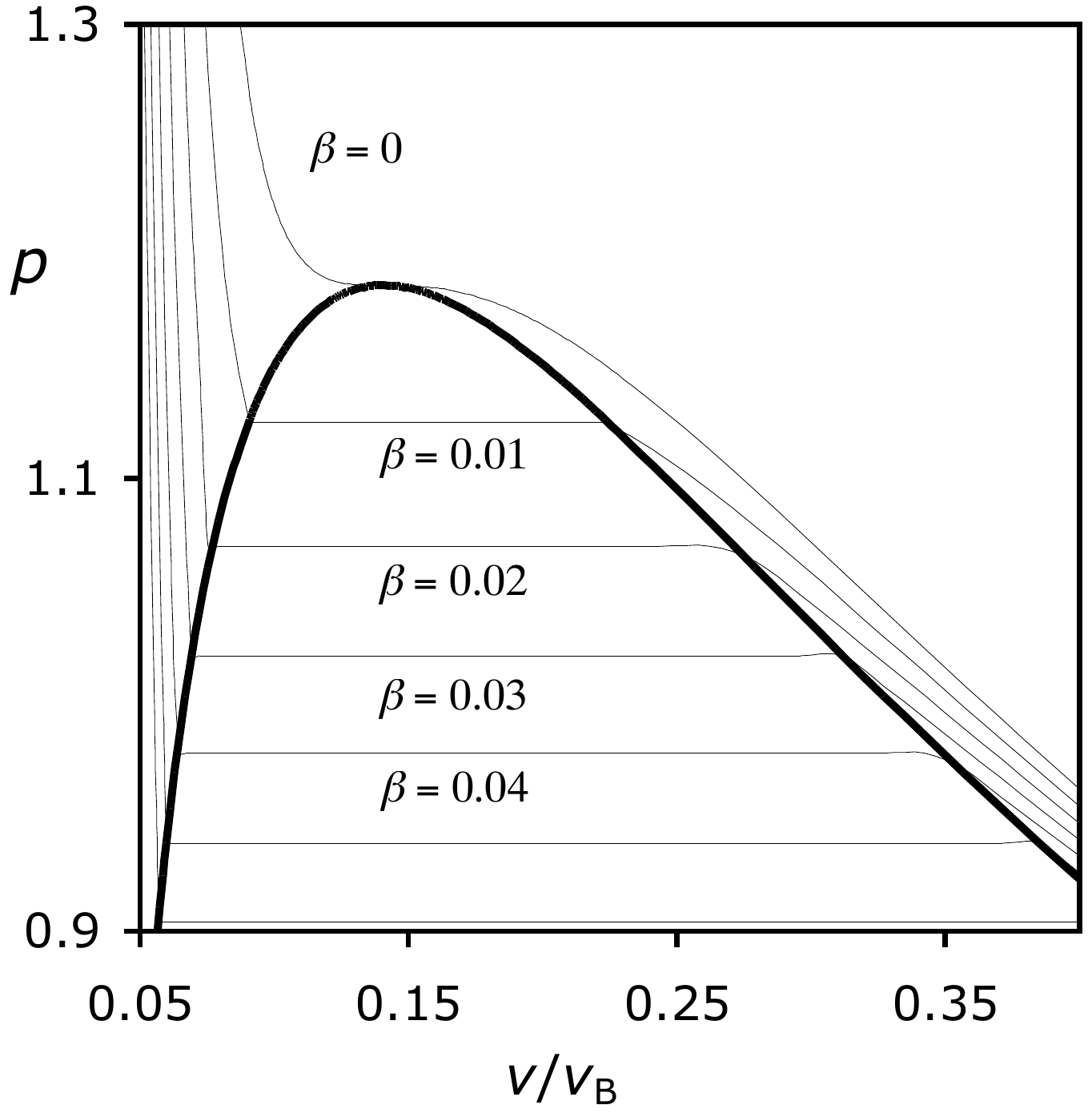}
\caption{
Coexistence diagram 
in dimensionless volume and
pressure for
a system
tuned by repulsion $\alpha$
to the monovalent ($\lambda=1$) critical point.
The divalent parameter $\beta$ rises from zero
upon addition of divalent ions, so that $\beta<0$ is not possible.
In similar coexistence diagrams for $0<\lambda<1$, the upper region will be
accessible.
}
\end{figure}

The dimensionless pressure ${p}(v) = P v_{\scriptscriptstyle B}/{k_{\text{\tiny B}}T}$ is
\begin{gather}
    {p_0}(v)
    +\frac{1}{2}\frac{v_{\scriptscriptstyle B}}{v}
    +
    \frac{1}{\beta}
    \bigg[
    \Big(1+ \beta\frac{v_{\scriptscriptstyle B}}{v}\Big)^{1/2} \!\! - 1
    \bigg]
    -\frac{1}{2}\Big(\frac{v_{\scriptscriptstyle B}}{v}\Big)^{3/2}
    \label{ptilde-of-v}
\end{gather}
where ${p_0} = P_0 v_{\scriptscriptstyle B}/{k_{\text{\tiny B}}T}$.
If the system is to be stable against collapse to $v=0$
due to the interaction term, we require
that ${p_0}(v)$ include a repulsion diverging faster than
$1/v^{3/2}$.  We will use the minimal choice
\begin{align}
    {p_0}(v)
    &\;=\;
    +\, \alpha (v_{\scriptscriptstyle B}/v)^2
    \label{minimal-choice}
\end{align}
(with $\alpha$ dimensionless).
The dimensionless electrical potential is
\begin{gather}
    \phi
    =
    \ln
    \frac{2}{\beta}
    \bigg[
    \Big(1 \!+\! \beta\frac{v_{\scriptscriptstyle B}}{v}\Big)^{1/2} \!\!\!\! - 1
    \bigg]
    -\frac{3}{2}
    \Big(\frac{v_{\scriptscriptstyle B}}{v}\Big)^{1/2}
    \!\!\!
    -\ln(v_{\scriptscriptstyle B}a )
    .
    \label{potential}
\end{gather}
From the Gibbs-Duhem relation (at constant $T$, or here $\beta$)
we obtain the chemical potential,
linking coexisting $v$ and $v'$
at specified $\beta$, as
$\mu(v) = {p}v -\int {p}(v) dv$, or
\begin{gather}
    \mu(v)
    \;=\;
    2\alpha/v
    -
    \tfrac{3}{2}
    v^{-1/2}
    -\ln \Big[v \!+\! \sqrt{v(v\!+\!\beta)}\Big]
    +\tfrac{1}{2}.
\end{gather}
Self-intersections of the curve
$(p(v),\mu(v))$
yield phase boundaries as
in Figs 1 and 2.

We solve our model in terms of the parameter
\begin{align}
    x &=
    \sqrt{v_{\scriptscriptstyle B}/v}
    .
\end{align}
At a critical point
the conditions
$p' (v) = p'' (v) = 0$ and
$p' (x) = p'' (x) = 0$ are equivalent.
Regarding the critical value of $p$ as a function of $x$ and $\beta$,
and introducing the parameter $\lambda = 1/\sqrt{1\!+\!\beta_c x_c^2}$,
we have the $x$-derivatives
\begin{align}
    {\tilde p}_{x}
    &\;=\;
    x_c\big[
    4\alpha \, x_c^2
    -\tfrac{3}{2}x_c
    +1
    \;
    +\lambda
    \big]
    \;=\;
    0
    \\
    {\tilde p}_{xx}
    &\;=\;
    12\alpha \, x_c^2
    -3x_c
    +1
    +\lambda^3
    \;=\;
    0
\end{align}
Eliminating $\alpha$ in favor of $\lambda$ gives
the line of critical points
\begin{align}
    p_c
    &\;=\; 
    x_c^2
    \big[
    \tfrac{1}{12} ( 1 - 6\lambda + \lambda^3 )
    + \lambda/(1+\lambda)
    \big]
    \label{pcrit}
    \\
    x_c
    &\;=\; 
    \tfrac{2}{3}
    [
    2
    +3\lambda
    -\lambda^3
    ]
    \label{xcrit}
    \\
    \alpha_c
    &\;=\;
    (1 +2\lambda -\lambda^3)/4x_c^2
    \label{alphacrit}
    \\
    \beta_c
    &\;=\;
    1/(x_c^2\lambda^2)
    \label{betacrit}
\end{align}
The interval $0 \le \lambda \le 1$ corresponds to $\infty\ge \beta \ge 0$.
Although we view $\alpha(\lambda)$ as the mechanism by which ionic critical
points can be tuned, we can also view
$\alpha_c = \alpha(\lambda)$
as a critical value of the repulsion strength.
With ionic conditions fixed,
we could drive transitions by modulating
$\alpha$, analogous to transitions in ionic
gels induced by variation of solvent composition \cite{Tanaka-1980}.

\begin{figure}
\includegraphics[width=1.0\columnwidth]
{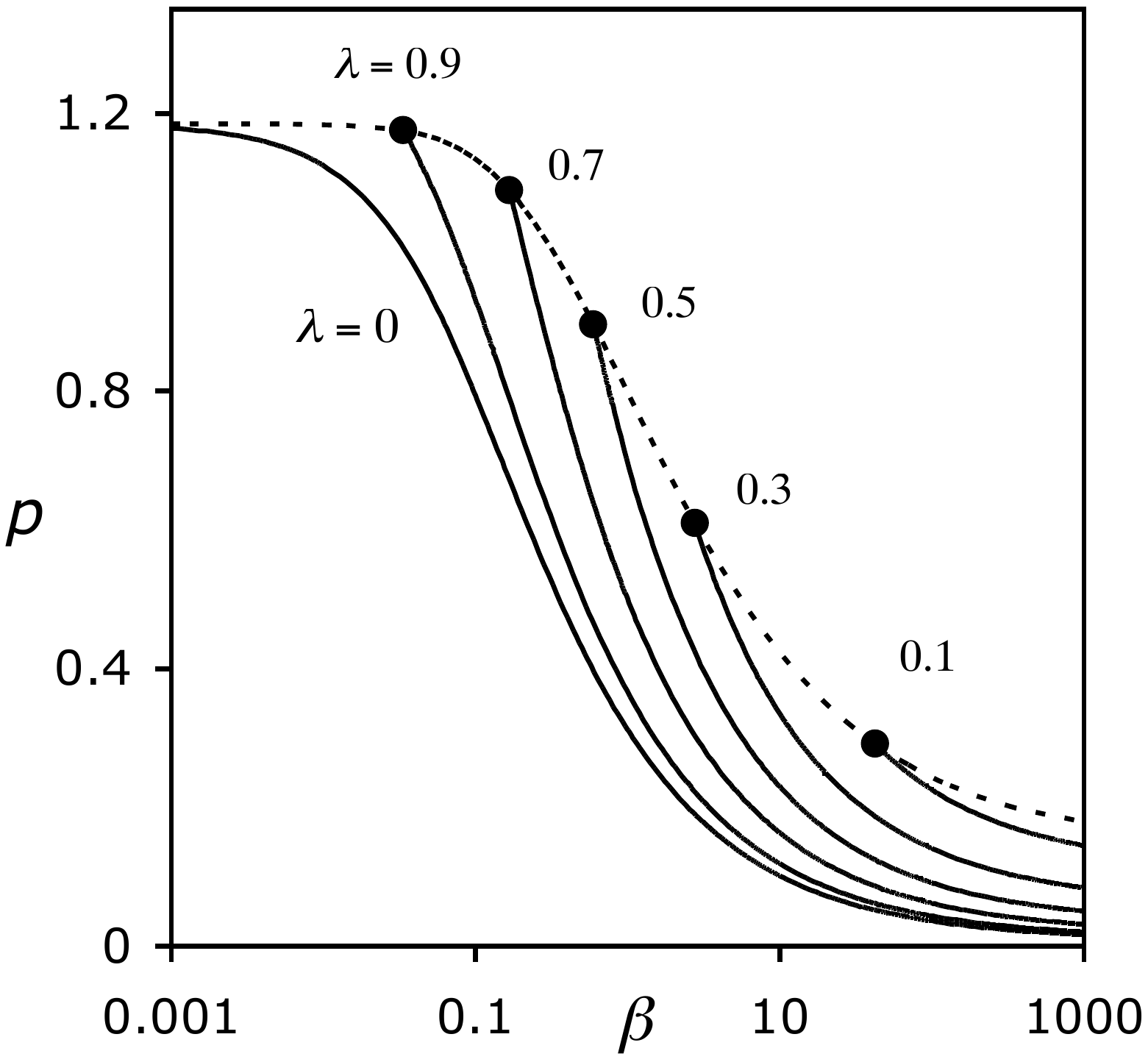}
\caption{
Critical line (dotted)
connecting the monovalent 
($\lambda=1$)
and divalent  ($\lambda=0$) critical points. 
The divalent parameter  is $\beta$ and
$p$ is dimensionless pressure.
Selected critical points and their 
phase boundaries (solid) are shown.
The phase boundary becomes inaccessible 
(moving to $\beta >\infty$) for the
critical point at $\lambda=0$. 
}
\end{figure}

About each critical point,
with $\delta v = v-v_c$ and $\delta \beta =\beta-\beta_c$,
we construct a Landau expansion for the pressure,
\begin{align}
    p(\beta,\epsilon)
    &\;=\;
    p_c -  {A}\delta \beta + {B}\delta \beta\,\epsilon
    - {C}\,\epsilon^3
\end{align}
where $\epsilon =  \delta v + \kappa \delta \beta$
is an order parameter linear in  $\delta v$ and $\delta \beta$.
In a potential application, we imagine tuning a system to 
a critical point, $\beta=\beta_c$ and $p=p_c$.
Within mean-field theory,
when the divalent ratio is changed by
$\delta\beta=\beta-\beta_c$
one predicts a singular expansion or contraction
\begin{align}
    \epsilon \approx
    -(A/C)^{1/3}
    \,
    (\delta\beta)^{1/3}
    \label{onethird}
\end{align}
where the cube root is taken with the same sign as $\delta\beta$.
As discussed below, this singular behavior in volume might in practice
be better monitored via the electric potential than the volume change
$\delta v$ or $\epsilon$.

To evaluate the Landau coeffients,
we expand the pressure around the critical point,
to third order in $\delta v=v-v_c$ and
to first order in $\delta\beta$, giving
\begin{gather}
    p(v,\beta)
    \;\approx\;
    p_c
    +{\tilde p}_{\beta}\delta \beta
    +{\tilde p}_{v\beta}\delta \beta \epsilon
    +\tfrac{1}{6} {\tilde p}_{vvv}
    \epsilon^3
\end{gather}
where $\beta$ and $v$ subscripts denote partial derivatives and
${\tilde p}$, ${\tilde p}_{v}$, etc.\ are critical values. We
have chosen $\kappa = {\tilde p}_{vv\beta} /{\tilde p}_{vvv}$
to eliminate an $\epsilon^2$ term. 
Evaluating the various derivatives,
the Landau parameters along the critical line are  
\begin{align}
    {A}
    &\;=\;
    \tfrac{1}{2} x_c^4\lambda^3 / (1+\lambda)^2
    ,\quad
    {B}
    \;=\;
    \tfrac{1}{4}x_c^6 \lambda^3
    \\
    {C}
    &\;=\;
    \tfrac{1}{48}x_c^8
    \big[
    2+12\lambda-7\lambda^3+3\lambda^5
    \big]
    \\
    \kappa
    &\;=\;
    (\lambda^3/x_c)(\lambda^2+1)/(8 x_c\alpha_c - \beta_c\lambda^5)
\end{align}
where $x_c$
and $\alpha_c$ are known functions of $\lambda$ from
Eqs (\ref{xcrit})
and (\ref{alphacrit}).
$C$ has no zeros within $0 \le \lambda \le 1$,
while $\kappa$ is zero only at $\lambda = 0$.
At the monovalent critical point
($\beta = 0$, i.e.\ $\lambda = 1$)
Eqs (\ref{pcrit}) through 
(\ref{betacrit}) yield
$p_c = 32/27 \approx 1.185$,
$x_c = 8/3 \approx 2.67$,
$\alpha_c  = 9/128 \approx 0.0703$,
the Landau coefficients are
\begin{align}
    {A}
    &\;=\;
    2^{9}/3^4
    \;\approx\;
    6.321
    \\
    {B}
    &\;=\;
    2^{16}/3^6
    \;\approx\; 89.90
    \\
    {C}
    &\;=\;
    5(2^{21}/3^8)
    \;\approx\;
    1598.2
\end{align}
and the order parameter is
$\epsilon = \delta v + (1/2)\delta\beta$.
Here the system enters coexistence for any value of $\beta>0$,
with $\epsilon \approx -(0.158)\,\beta^{1/3}$. 
A system tuned to the monovalent critical point 
moves with infinite response towards a smaller volume with the 
introduction of any divalents.
In the liquid-gas analogy, this path follows the critical isobar.

At the purely divalent critical point ($\beta = \infty$, $\lambda = 0$) 
both ${A}$ and ${B}$ vanish, but
we can here characterize the line of critical points by
the alternative parameterization
\begin{align}
    p(\beta,\epsilon)
    &\;=\;
    p_c 
    +{A'} \delta\big(\beta^{\scriptscriptstyle -1/2}\big)
    -{B'} \delta\big(\beta^{\scriptscriptstyle -1/2}\big)
    - {C}\,\epsilon^3
    \\
    {A'}
    &\;=\;
    x_c(1-\lambda)^{3/2}/(1+\lambda)^2
    \\
    {B'}
    &\;=\;
    \tfrac{1}{2}x_c^3
    (1-\lambda^2)^{3/2}
\end{align}
The parameter $\beta^{\scriptscriptstyle -1/2}$
is proportional to the monovalent
concentration at fixed divalent concentration.
The volume singularity (\ref{onethird}) can be rewritten as
\begin{align}
    \epsilon \approx
    +\Big[
    \frac{{A'}}{{C}}
    \,\delta\big(\beta^{\scriptscriptstyle -1/2}\big)
    \Big]^{1/3}
    \label{onethirdalt}
\end{align}
where again the cube root has the same sign as its argument.
At the divalent critical point ($\beta = \infty$, $\lambda = 0$)
Eqs (\ref{pcrit}) through 
(\ref{betacrit}) yield
$p_c = 4/27 \approx 0.1481$,
$x_c = 4/3$, $\alpha_c = 9/64 \approx 0.1406$,
\begin{align}
    {A'}
    &=
    4/3
    ,
    \quad
    {B'}
    =
    32/27
    \;\approx\;
    1.185
    \\
    {C}
    &\;=\;
    2^{13}/3^9
    \;\approx\; 0.4162
\end{align}
For this divalent critical point at $\beta_c^{\scriptscriptstyle -1/2}=0$, 
the order parameter is simply the volume,
$\epsilon = \delta v$.
A purely divalent system that is tuned to be critical
moves with infinite response towards a larger volume
with the introduction of any monovalents, as
$\delta v \approx (1.214)\,\beta^{-1/6}$.

Returning to the dimensionless potential $\phi = q_0\Phi/{k_{\text{\tiny B}}T}$ in
Eq (\ref{potential}),
along the critical line with $a$ constant
we find
\begin{align}
    \Big[
    \frac{\partial\phi}{\partial v}
    \Big]_c
    &\;=\;
    \frac{3}{4}
    \frac{x_c^3}{v_{\scriptscriptstyle B}}
    \Bigg[
    \frac{1+2\lambda - \lambda^3}
    {2+3\lambda - \lambda^3}
    \Bigg]
\end{align}
Since $\delta \phi\approx \phi^c_v \delta v$,
the potential will display the same power-law singularity
as the volume.

Our simplified framework is intended to motivate investigation of ionic phase
transitions in ever simpler systems.  As a fascinating analogue of liquid-gas
transitions, and in the context of ionic gels, such transitions have been
a topic of discussion for many years.  Possible roles in biological
systems \cite{Tasaki-Verdugo} and potential applications such as metal ion
detection and electrical sensitivity \cite{Tanaka-1982,Jackson-1997} should remain
applicable to phase transitions as discussed here.

\begin{acknowledgments}
The authors would like to thank colleagues at the Department of Physics 
and Astronomy at Washington State University for their support and helpful
conversations.
\end{acknowledgments}

\end{document}